\documentclass[reprint,aps,prl,superscriptaddress]{revtex4-2}

\usepackage{graphicx}% Include figure files
\usepackage{dcolumn}% Align table columns on decimal point
\usepackage{bm}% bold math
\usepackage{SIunits}	% Pour utiliser les unités physique
\usepackage{color}
\usepackage{booktabs}

\definecolor{r}{rgb}{0.86,0.08,0.23}

\begin{document}

\title{Broad diversity of near-infrared single-photon emitters in silicon}

\author{A. Durand} 	\affiliation{Laboratoire Charles Coulomb, Universit\'e de Montpellier and CNRS, 34095 Montpellier, France} 
\author{Y. Baron}		\affiliation{Laboratoire Charles Coulomb, Universit\'e de Montpellier and CNRS, 34095 Montpellier, France} 
\author{W. Redjem}		\affiliation{Laboratoire Charles Coulomb, Universit\'e de Montpellier and CNRS, 34095 Montpellier, France} 
\author{T. Herzig}		\affiliation{Division of Applied Quantum Systems, Felix-Bloch Institute for Solid-State Physics, University Leipzig, Linn\'eestra\ss e 5, 04103 Leipzig, Germany} 
\author{A. Benali}		\affiliation{CNRS, Aix-Marseille Universit\'e, Centrale Marseille, IM2NP, UMR 7334, Campus de St. J\'er\^ome, 13397 Marseille, France} 
\author{S. Pezzagna}	\affiliation{Division of Applied Quantum Systems, Felix-Bloch Institute for Solid-State Physics, University Leipzig, Linn\'eestra\ss e 5, 04103 Leipzig, Germany} 
\author{J. Meijer}		\affiliation{Division of Applied Quantum Systems, Felix-Bloch Institute for Solid-State Physics, University Leipzig, Linn\'eestra\ss e 5, 04103 Leipzig, Germany} 
\author{A.~Yu.~Kuznetsov}\affiliation{Department of Physics, University of Oslo, NO-0316 Oslo, Norway} 
\author{J.-M. G\'erard}	\affiliation{Department of Physics, IRIG-PHELIQS, Univ. Grenoble Alpes and CEA, F-38000 Grenoble, France} 
\author{I. Robert-Philip}    \affiliation{Laboratoire Charles Coulomb, Universit\'e de Montpellier and CNRS, 34095 Montpellier, France} 
\author{M. Abbarchi}	 \affiliation{CNRS, Aix-Marseille Universit\'e, Centrale Marseille, IM2NP, UMR 7334, Campus de St. J\'er\^ome, 13397 Marseille, France} 
\author{V. Jacques}	 \affiliation{Laboratoire Charles Coulomb, Universit\'e de Montpellier and CNRS, 34095 Montpellier, France} 
\author{G. Cassabois}	 \affiliation{Laboratoire Charles Coulomb, Universit\'e de Montpellier and CNRS, 34095 Montpellier, France} 
\author{A. Dr\'eau}	\email{anais.dreau@umontpellier.fr}	 \affiliation{Laboratoire Charles Coulomb, Universit\'e de Montpellier and CNRS, 34095 Montpellier, France} \email{anais.dreau@umontpellier.fr}

\begin{abstract}
We report the detection of individual emitters in silicon belonging to seven different families of optically-active point defects. 
These fluorescent centers are created by carbon implantation of a commercial silicon-on-insulator wafer usually employed for integrated photonics. 
Single photon emission is demonstrated over the [1.1,1.55]-$\mu$m range, spanning the O- and C-telecom bands. 
We analyse their photoluminescence spectrum, dipolar emission and optical relaxation dynamics at 10K. 
For a specific family, we show a constant emission intensity at saturation from 10K to temperatures well above the 77K-liquid nitrogen temperature. 
Given the advanced control over nanofabrication and integration in silicon, these novel artificial atoms are promising candidates for Si-based quantum technologies. 

\end{abstract}

\maketitle

The boom of silicon in semiconductor technologies was closely tied to the ability to control its density of lattice defects~\cite{yoshida_defects_2015}. 
After being regarded as detrimental to the crystal quality in the first half of the 20$^{\mathrm{th}}$ century~\cite{queisser_defects_1998}, point defects have become an essential tool to tune the electrical properties of this semiconductor, leading to the development of a flourishing silicon industry~\cite{yoshida_defects_2015}.  
At the turn of the 21$^{\mathrm{st}}$ century, progress in Si-fabrication and implantation processes has triggered a radical change by enabling the control of these defects at the single level~\cite{morello_single-shot_2010}. 
This paradigm shift has brought silicon into the quantum age, where individual dopants are nowadays used as robust quantum bits to encode and process quantum information~\cite{he_two-qubit_2019}. 
These individual qubits can be efficiently controlled and detected by all-electrical means~\cite{he_two-qubit_2019}, but have the drawback of either being weakly coupled to light~\cite{steger_quantum_2012} or emitting in the mid-infrared range~\cite{morse_photonic_2017} unsuitable for optical fiber propagation.
In order to isolate matter qubits that feature an optical interface enabling long-distance exchange of quantum information while benefiting from well-advanced silicon integrated photonics~\cite{silverstone_silicon_2016}, one strategy is to investigate defects in silicon that are optically-active in the near-infrared telecom bands~\cite{bergeron_silicon-integrated_2020, weiss_erbium_2020}.

Point defects emitting light exist in various semiconductors~\cite{weber_quantum_2010, aharonovich_solid-state_2016, zhang_material_2020}.~While individual solid-state artificial atoms have been observed in several wide bandgap semiconductors such as diamond~\cite{gruber_scanning_1997, bradac_quantum_2019}, silicon carbide~\cite{christle_isolated_2015,widmann_coherent_2015} or hexagonal boron nitride~\cite{tran_quantum_2016}, silicon is lagging behind~\cite{zhang_material_2020}.
The first optical detection of a single optically-active defect in silicon has only been reported recently~\cite{redjem_single_2020} and reproduced in~\cite{hollenbach_engineering_2020}.
Besides this defect related to a carbon-complex called the G-center~\cite{davies_optical_1989,beaufils_optical_2018,chartrand_highly_2018}, we report here that silicon, despite its small bandgap, hosts a large variety of emitters that can be optically isolated at single scale. 
We identify six new families of individual fluorescing defects in carbon-implanted silicon, that emit single photons in the near-infrared range covering the O- and C-telecom bands. 
We investigate their photoluminescence spectra, dipolar emission and radiative recombination dynamics through lifetime measurements. 
At last, we demonstrate that G-centers have a constant emission intensity at saturation, from 10K to well above liquid-nitrogen temperatures.

\begin{figure*}
  \includegraphics[width=\textwidth]{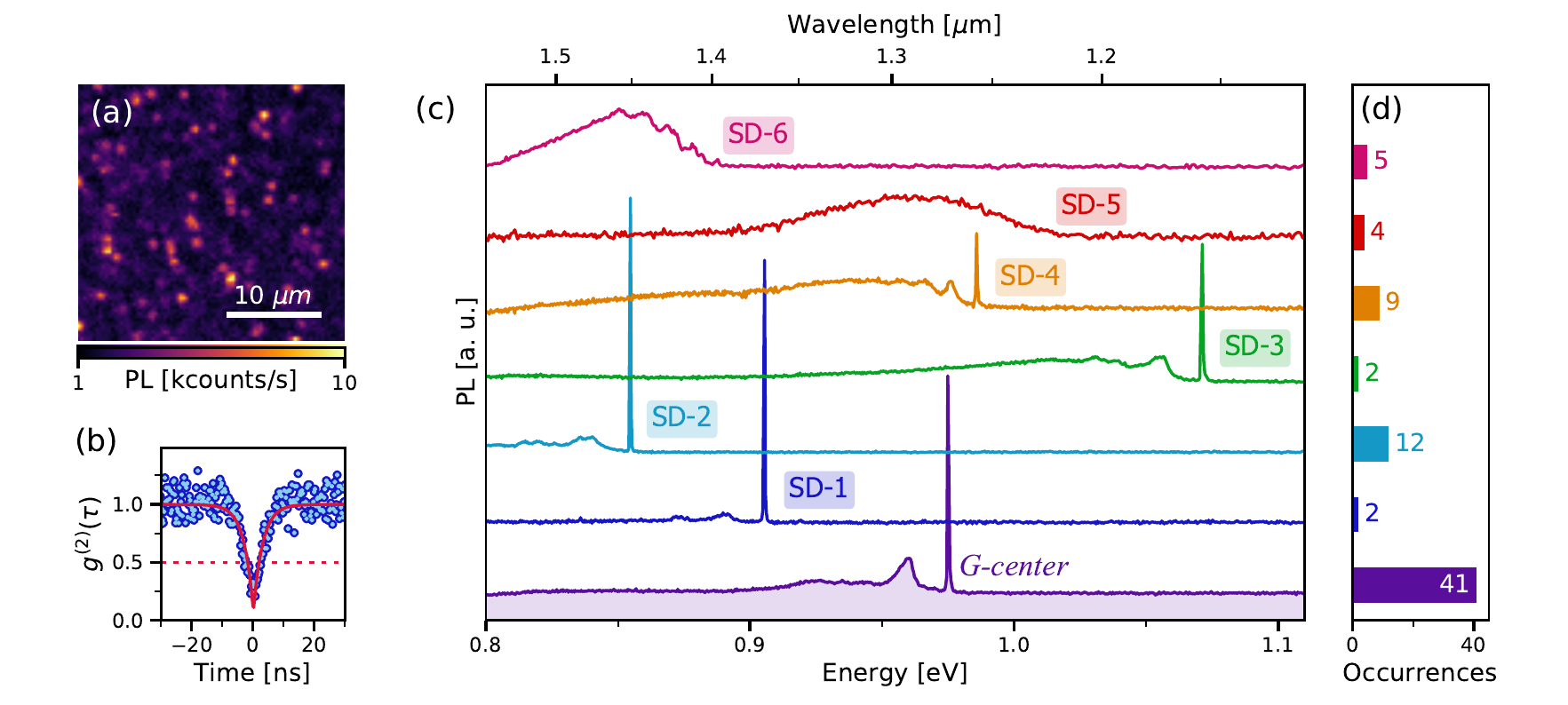}
  \caption{
		(a) PL raster scan recorded at \unit{10}{\kelvin} under excitation of the carbon-implanted SOI sample with a 532-nm laser at  $ \unit{10}{\mu\watt}$.
		Each isolated bright spot is a single emitter. 
		(b) Second-order autocorrelation function $g^{2}(\tau)$ measured on a defect belonging to family SD-1. 
		At zero-delay, the curve displays a strong antibunching below the single-emitter threshold :  $g^{(2)}(0) < 0.5$. 
		There is no background correction. Data are fitted (solid line) with a two-level model~\cite{beveratos_bunching_2002}.
		(c) PL spectra recorded on seven different individual single-photon emitters. 
		The bottom spectrum is associated with the G-center in silicon, recently isolated at single scale~\cite{redjem_single_2020}. 
		Spectra SD-1 to SD-6 correspond to six new families of single emitters randomly distributed over the sample. 
		(d) Histogram of the number of defects per family. 
  \label{fig:first}
}
\end{figure*}

The investigated sample comes from a commercial silicon-on-insulator (SOI) wafer (Soitec). 
The top 220-nm thick silicon layer was implanted with 36-keV carbon ions at a fluence of \unit{5 \times 10^{13}}{\centi\meter^{-2}}. 
To heal the lattice from damages generated during this process, the sample underwent a subsequent flash annealing at 1050$^{\circ}$C during 20 seconds. 
The experimental setup consists in a low-temperature scanning confocal microscope operating with above-bandgap optical excitation at 532 nm. 
The sample photoluminescence (PL) is collected by a high numerical aperture microscope objective (NA=0.85) and measured with fiber-coupled single-photon detectors featuring a detection efficiency of $10 \%$.
The optical detection window was set to cover the near-infrared range from 1.1 $\mu$m to 1.55 $\mu$m.
More details about the sample preparation and optical setup can be found in~\cite{redjem_single_2020}.

\begin{table*}
\centering
\renewcommand*{\arraystretch}{1.2}
 \caption{Summary of the optical properties for the different families of single defects.
ZPL uncertainties for families G, SD-2 and SD-4 correspond to the standard deviation calculated over the full family set.
Statistical data for G-centers are taken from ref.~\cite{redjem_single_2020}.
 } 
 \label{tab:family_recap}
\begin{tabular}{ @{}l >{\centering} m{5em}>{\centering} m{5em}>{\centering} m{5em}>{\centering} m{5em}>{\centering} m{5em}>{\centering}  m{5em} c @{}}\toprule
Family  & SD-1 & SD-2 & SD-3 & SD-4 & SD-5 & SD-6& G \\ \midrule
Spectrum with ZPL 		 & Yes & Yes & Yes & Yes & No & No & Yes\\
ZPL energy [meV]		 &  $\simeq 905$ & 856 $\pm$ 3 & $\simeq 1071$  & 989 $\pm$ 6 & - & - & 977 $\pm$ 7 \\
ZPL wavelength [nm]	  &  $\simeq 1369$  & 1448 $\pm$ 5 & $\simeq 1157$ & 1253 $\pm$ 7 & - & - & 1269 $\pm$ 9\\
Debye-Waller factor [\%]  & 35 & 25 & 3 & 2 & - & - & 15\\
1$^{\mathrm{st}}$ phonon replica energy [\milli e\volt]   & 14.5 & 14.5 & 14.5 & 9.5 & - & - & 14.5\\
PL intensity at saturation [\kilo c\per\second]  & 13 & 9 & 8 & 14 & 22 & 10& 16 \\
Resistant to thermal cycles  & Yes & Yes & Yes & Yes & No & Yes & Yes\\ 
Number of emission dipoles & 1 & 1 & 1 & 1 & 1 & 1 & 1\\
ES lifetime(s) [\nano\second]   & 14.4 $\pm$ 0.4 & 30.6 $\pm$ 0.5 & 30 $\pm$ 1 &  26 $\pm$ 1 & 4.4 $\pm$ 0.7 & 6.7 $\pm$ 0.8 & 36 $\pm$ 4\\
 & & & & & 19 $\pm$ 5  &  35 $\pm$ 4&  \\
\bottomrule
\end{tabular}
\end{table*}

A PL image of the sample at 10K reveals a multitude of fluorescing spots (Fig.~\ref{fig:first}(a)).
Analyzing their photon emission statistics with a Hanbury-Brown and Twiss setup indicates that all the isolated bright spots appear to be single emitters. 
Their autocorrelation function $g^{(2)}(\tau)$ presents a clear antibunching that fulfills the single-emitter condition: $g^{(2)}(0) < 0.5$~\cite{eisaman_invited_2011}, as displayed on Fig.~\ref{fig:first}(b) (see SI). 
By examining the PL spectra of these isolated bright spots (Fig.~\ref{fig:first}(c)), we observe a broad diversity of emission between 1.1 and 1.55 $\mu$m associated with different fluorescing defects in the C-implanted silicon. 
Besides individual G-centers investigated in a previous study~\cite{redjem_single_2020}, we identify six new families of single emitters that we label from SD-1 to SD-6.
Apart from the SD-2 family that might be related to interstitial carbon defects $C(i)$~\cite{thonke_new_1987}, these novel individual single-photon emitters could not be here related to luminescent defects in silicon previously identified in the literature through optical spectroscopy on large ensembles~\cite{davies_optical_1989}. 
Given the small occurrence of those families with respect to the G-centers (Fig. \ref{fig:first}(d)), one explanation could be that their PL signal detected here at single scale is hidden by the emission of G-centers and possibly other defects in ensemble measurements.

\begin{figure}[h]
  \includegraphics[width=\columnwidth]{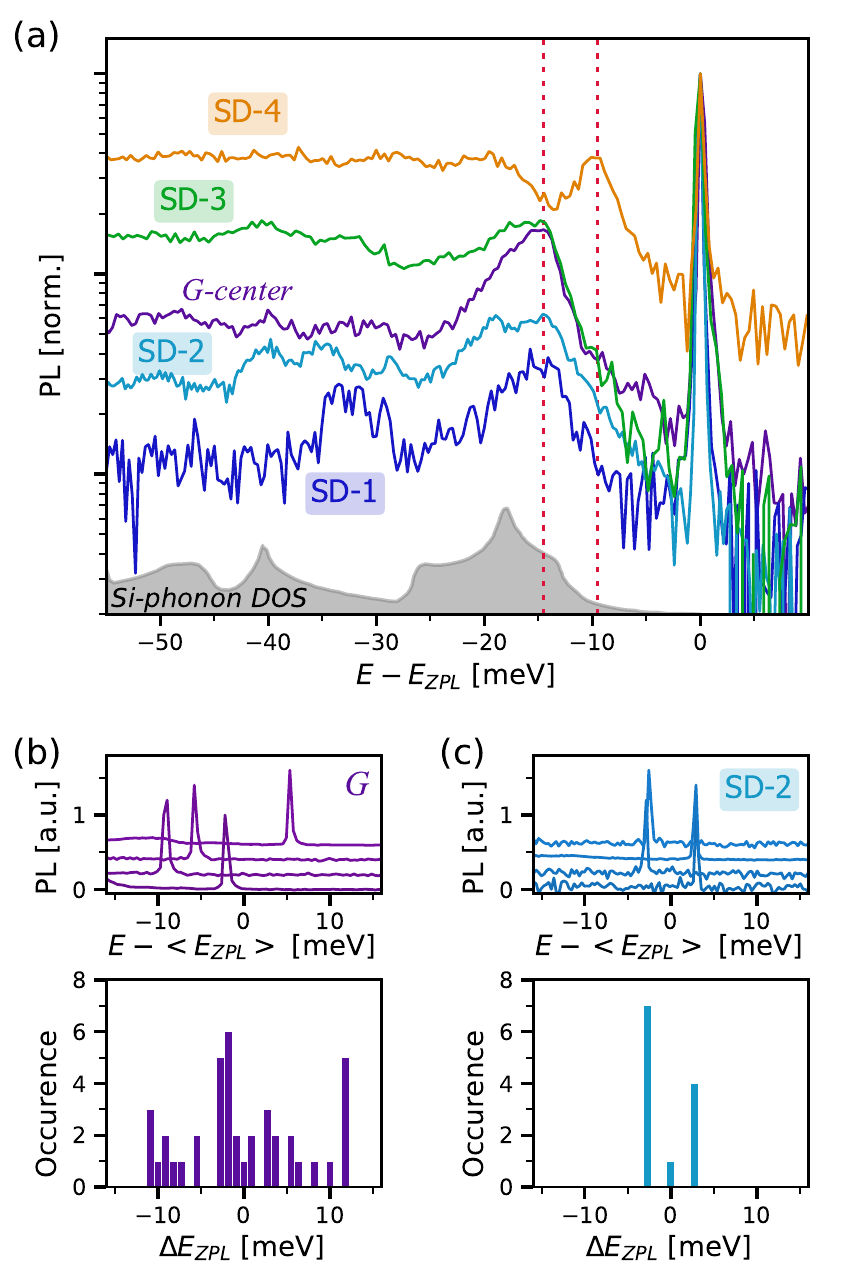}
  \caption{
	(a) Comparison of the phonon-sideband for the defects with a ZPL. 
	Spectra are normalized to the ZPL maximum and plotted with respect to their ZPL energy $E_{ZPL}$. 
	The vertical lines show the position of the first phonon replica at \unit{9.5}{\milli e\volt} and \unit{14.5}{\milli e\volt}.
	The gray-shaded area indicates the silicon-phonon density of states. 
	(b-c) (top) Typical PL spectra measured on individual defects for families G and SD-2 respectively. 
	Spectra are plotted with respect to the mean ZPL energy position $\langle E_{ZPL} \rangle$. 
	(bottom) Distribution of the ZPL energy shift compare to the middle energy.
  \label{fig:spectra_comparison}
  }
\end{figure}

These novel individual defects in silicon all emit in the near-infrared region. 
Their PL can even match the O- to C-telecom bands that covers the range from 1.26 to 1.56 $\mu$m.
We first observe that the PL spectrum strongly differs from one type of defect to another. 
While the zero-phonon-line (ZPL) dominates over the vibronic spectrum for families SD-1 and SD-2, it becomes less and less intense for families SD-3 and SD-4, until not being detectable for families SD-5 and SD-6 (Fig.~\ref{fig:first}(c)). 
The Debye-Waller factor (DW) meters the proportion of photons emitted in the ZPL and provides information about the electron-phonon coupling~\cite{dreyer_first-principles_2018}. 
As indicated in Table \ref{tab:family_recap}, it reaches values as high as 35\% and 25 \% for SD-1 and SD-2 defects respectively,  whereas it is limited to 2-3 \% for families SD-3 and SD-4. 
A closer inspection of the phonon-sideband (PSB) reveals that the first phonon replica appears either at 9.5 meV (SD-4) or at 14.5 meV (SD-1 to SD-3, G-center), as depicted on Figure \ref{fig:spectra_comparison}(a) and Table \ref{tab:family_recap}.
We note that the PSB of family SD-6 exhibits a periodicity of 9.5 meV. 
Since these energies do not correspond to any of the maxima of the silicon-phonon density of states (Fig.~\ref{fig:spectra_comparison}(a)), the vibronic spectrum likely results from phonons combining localized and Bloch vibrational states~\cite{estreicher_first-principles_2003}.

Comparing the ZPL position for the most prevalent families SD-2 and G shows two radically different distributions.
The ZPL position randomly spreads over 20 meV around 977 meV for the G-centers due to local strain variations~\cite{tkachev_piezospectroscopic_1978} (Fig.~\ref{fig:spectra_comparison} (b)).~By contrast, the ZPL of SD-2 defects is always found at three different energies equally split by 3 meV around 856 meV(Fig.~\ref{fig:spectra_comparison}(c)).
Such a behavior suggests three defect configurations for SD-2 defects, which are less sensitive to local strain fluctuations, an appealing property in the prospects of indistinguishable single-photon emission~\cite{sipahigil_indistinguishable_2014}.
We note that the average ZPL position for SD-2 centers matches the 856-meV value reported for interstitial carbon defects~\cite{thonke_new_1987}.
Advanced theoretical calculations would be required to explain the observed ZPL distribution.

\begin{figure}[h]
  \includegraphics[width=\columnwidth]{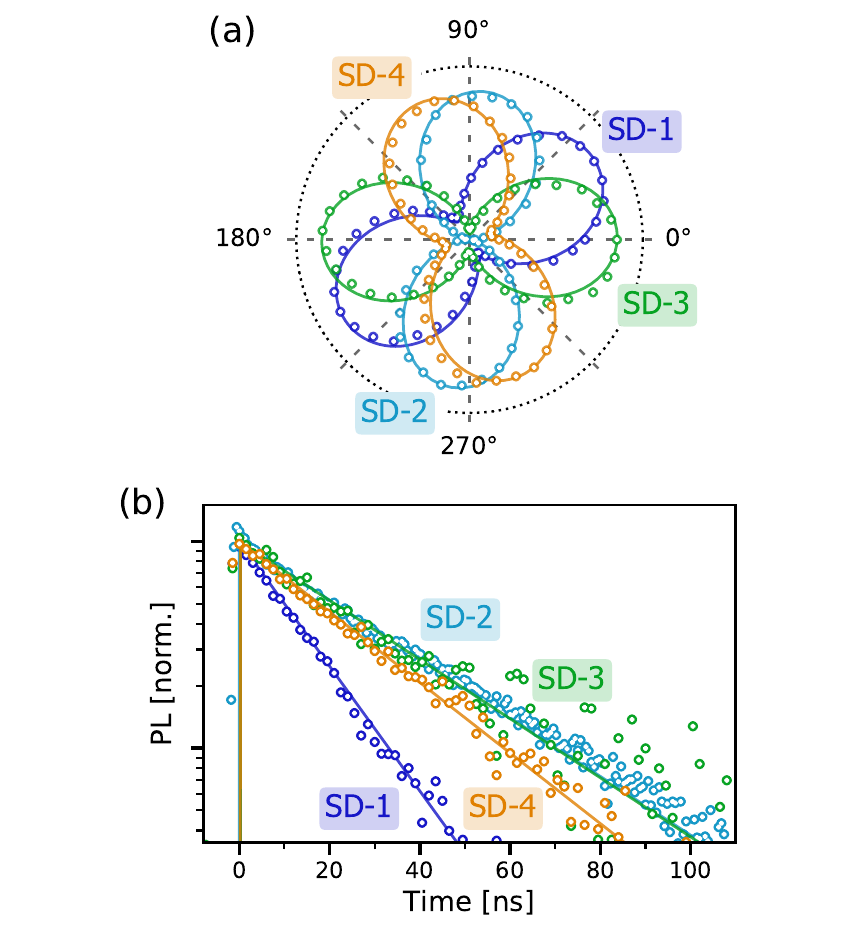}
  \caption{(a) Emission polarization diagram measured on single defects from families SD-1 to SD-4.
  The PL signal is recorded while rotating a polarizer in the detection path and corrected from background counts. 
  The $0^{\circ}$ and $90^\circ$ directions match the crystal axes $[110]$ and $[1\bar{1}0]$.
  Solid lines are fits using a $\cos^2(\theta)$ function.
  (b) Time-resolved PL decay recorded on the same defects under 50-ps pulse excitation at 532 nm. 
  The excited-state lifetime is extracted by fitting the data with a single exponential function (solid line). 
  Data related to SD-5 and  SD-6 defects can be found in SI. 
  \label{fig:dipole_lifetime}
  }
\end{figure}

Besides being near-infrared single-photon emitters, these single defects in silicon share additional interesting PL properties. 
First, these individual emitters are perfectly photostable, with neither blinking nor bleaching observed while recording their PL signals over time. 
Furthermore, no center fades away after room-temperature warming-up followed by cooling-down to 10K, except the defects of family SD-5 that either disappear or appear in optical scans (see SI). 
Then, like individual G-centers in silicon~\cite{redjem_single_2020}, these novel single-photon emitters show a bright emission under green optical excitation, with PL intensity at saturation on the order of 10-20 kcounts/s (Table \ref{tab:family_recap}). 
These PL count rates are high considering the poor quantum efficiency of our detectors (only 10\%) and that we collect at best 2\% of the emitted photons due to the high refractive index of silicon ($n\sim 3.5$)~\cite{redjem_single_2020}. 
At last, all defect families emit linearly-polarized single photons. 
As shown on Figure \ref{fig:dipole_lifetime}(a) for families SD-1 to SD-4, each of the PL polarization diagrams measured on individual centers shows the characteristic emission of a single dipole~\cite{elliott_polarization_1958} (see SI for families SD-5 and SD-6). 
Although the interstitial Si atom inside the G-centers as well as interstitial carbons are reported to be mobile respectively above 30K~\cite{odonnell_origin_1983} and 77K~\cite{davies_optical_1989}, we did not observe any change in the emission polarization diagram of G- and SD-2 defects neither at 130K and 70K respectively nor after successive thermal cycles of the cryostat (see SI).

To characterize the relaxation dynamics of the individual single-photon emitters, we performed time-resolved PL measurements under 532-nm pulse excitation. 
The photon histograms recorded on single defects belonging to families SD-1 to SD-4 are shown on Figure 3(b). 
For those four families, we observe a mono-exponential decay providing a single excited-state (ES) lifetime ranging between 14 ns to 30 ns (Table \ref{tab:family_recap}).
On the contrary, families SD-5 and SD-6 feature a bi-exponential decay, with a short lifetime roughly at 5 ns and a long one around 19 ns and 35 ns respectively (see Table \ref{tab:family_recap} \& SI). 
All these values are orders of magnitude shorter than the ones measured on emitter ensembles in silicon, such as erbium dopants~\cite{weiss_erbium_2020} or T-centers~\cite{bergeron_silicon-integrated_2020}.
Consequently, these defects are already advantageous to develop bright Si-based single-photon sources even without requiring PL enhancement by Purcell-effect~\cite{purcell_resonance_1946}.

\begin{figure}
  \includegraphics[width=\columnwidth]{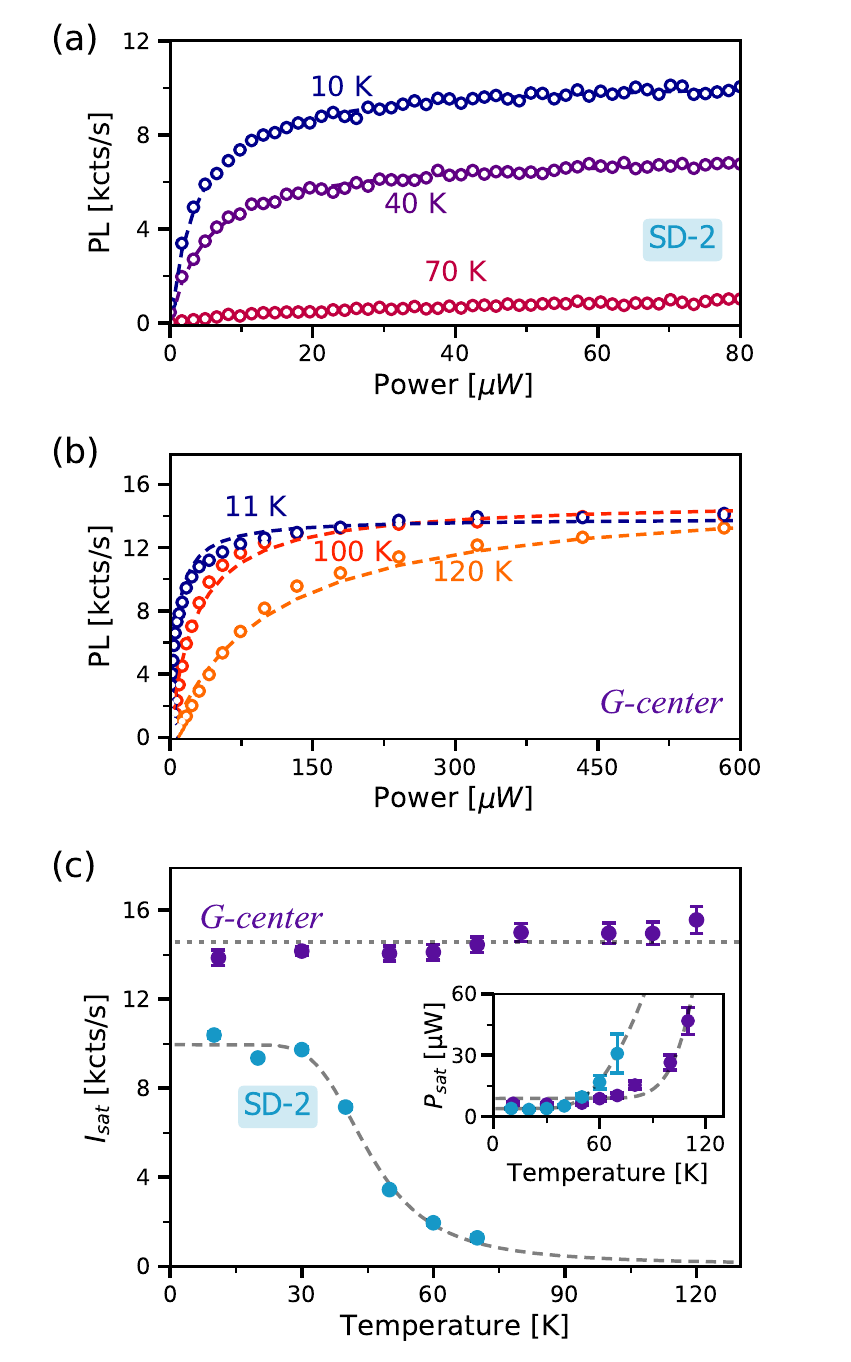}
  \caption{
	(a-b) Background-corrected PL saturation curves measured respectively on a single G-defect and a single SD-2 defect for different temperatures.
	Data is fitted with Eq. \ref{eq:saturation_curve} to extract the saturation power $P_{sat}$ and intensity at saturation $I_{sat}$.
	(c) Corresponding evolution of $I_{sat}$ and $P_{sat}$ (inset) with temperature for the SD-2 (blue) and G- (purple) defects. 
	The dash lines are data fitting (see main text for $P_{sat}$ data).
	The $I_{sat}$ data for the SD-2 defect are fairly reproduced by the function: $a'/(1 + b' \exp(-E_{a2}/(k_B T))$, where $a'$ and $b'$ are free parameters, and $E_{a2} = 24 \pm 3$ meV the activation energy.   
	The dotted line for the G-data is a guide for the eye.
  \label{fig:both_temperature}
}
\end{figure}

We finally analyze the PL saturation curves as a function of temperature. 
Here we focus on the most frequent families SD-2 and G (Fig. \ref{fig:first}(d)). 
Figures \ref{fig:both_temperature} (a-b) show the PL intensity evolution with optical power $P$ at different temperatures for two single SD-2 and G-defects. 
Data follow the behavior from a two-level system modeled by the standard saturation equation:
\begin{equation}
I(P)= I_{sat} \frac{P} {P + P_{sat}}
\label{eq:saturation_curve}
\end{equation}
with $I_{sat}$ being the maximum PL intensity at saturation and $P_{sat}$ the saturation power. 
We observe for both defects a rise of $P_{sat}$ with increasing temperature  (Fig.~\ref{fig:both_temperature}(c) Inset). 
This behavior stems from the thermal activation of non-radiative decay channels, as evidenced by the decrease of ES lifetimes at elevated temperatures~\cite{beaufils_optical_2018} (see SI). 
The evolution of $P_{sat}$ with temperature $T$ is fitted with: ${a+ b \exp(-E_{a1}/(k_BT)})$, with $a$ and $b$ as free constants, $k_B$ the Boltzmann constant and $E_{a1}$ the activation energy that gives the respective values $E_{a1}^{(SD-2)} = 25 \pm 1 $ meV and $E_{a1}^{(G)} = 95 \pm 9$ meV for SD-2 and G-defects. 
The most striking feature is related to the PL intensity at saturation: while $I_{sat}$ drops quickly above 30K for the family SD-2, it stays constant for G-centers well-above the 77K-liquid nitrogen temperature (Fig.~\ref{fig:both_temperature}).
Since the PL counts mirror the ES population of the emitters, this indicates that the G-defects behave as a closed-system when the temperature raises whereas the SD-2 defects are coupled to their environment.

In conclusion, we report the isolation at single scale of new optically-active point defects in silicon.~These novel individual emitters provide a wide diversity of bright, linearly-polarized single-photon emission in the near-infrared range, matching the O- and C-telecom bands.~We further demonstrate that some single defects exhibit additional appealing properties such as a small spread of the ZPL energies or a strong PL intensity well above the liquid-nitrogen temperature. 

This multitude of fluorescent artificial atoms available in silicon could open a new path in exploring Si-based quantum technologies.~These single defects could serve as building blocks to develop efficient and deterministic sources of photonic qubits in a material widely used for integrated photonics applications~\cite{silverstone_silicon_2016, qiang_large-scale_2018}.  
Combining optical and microwave magnetic excitations could enable to investigate the spin properties attached to these unidentified new single-photon emitters, in view of isolating individual spin-photon interfaces in silicon operating at telecom wavelength~\cite{bergeron_silicon-integrated_2020}.~The advance nanotechnology based on this semiconductor, combined with the myriad of available defects~\cite{davies_optical_1989,pichler_intrinsic_2004} points the way towards thriving quantum applications based on single defects in silicon~\cite{koenraad_single_2011,weber_quantum_2010}.

\bibliographystyle{ieeetr}

\bibliography{biblio}

\end{document}